\documentclass[Physsubmission, Phys]{SciPost}

\usepackage{amsmath}
\usepackage{amsfonts}

\hypersetup{
    colorlinks,
    linkcolor={red!50!black},
    citecolor={blue!50!black},
    urlcolor={blue!80!black}
}

\DeclareSymbolFont{usualmathcal}{OMS}{cmsy}{m}{n}
\DeclareSymbolFontAlphabet{\mathcal}{usualmathcal}

\begin{document}

\begin{center}{\Large \textbf{
Impact parameter dependence of color charge correlations in the proton\\
}}\end{center}

\begin{center}
A. Dumitru\textsuperscript{1,2$\star$},
H. M\"antysaari\textsuperscript{3,4} and
R. Paatelainen\textsuperscript{4}
\end{center}

\begin{center}
{\bf 1} Department of Natural Sciences, Baruch College, CUNY, 17 Lexington Avenue, New York, NY 10010, USA
\\
{\bf 2} The Graduate School and University Center, The City University of New York, 365 Fifth Avenue, New York, NY 10016, USA
\\
{\bf 3} Department of Physics, University of Jyväskylä,  P.O. Box 35, 40014 University of Jyväskylä, Finland
\\
{\bf 4} Helsinki Institute of Physics, P.O. Box 64, 00014 University of Helsinki, Finland
\\

* adrian.dumitru@baruch.cuny.edu
\end{center}

\begin{center}
\today
\end{center}


\definecolor{palegray}{gray}{0.95}
\begin{center}
\colorbox{palegray}{
  \begin{tabular}{rr}
  \begin{minipage}{0.1\textwidth}
    \includegraphics[width=22mm]{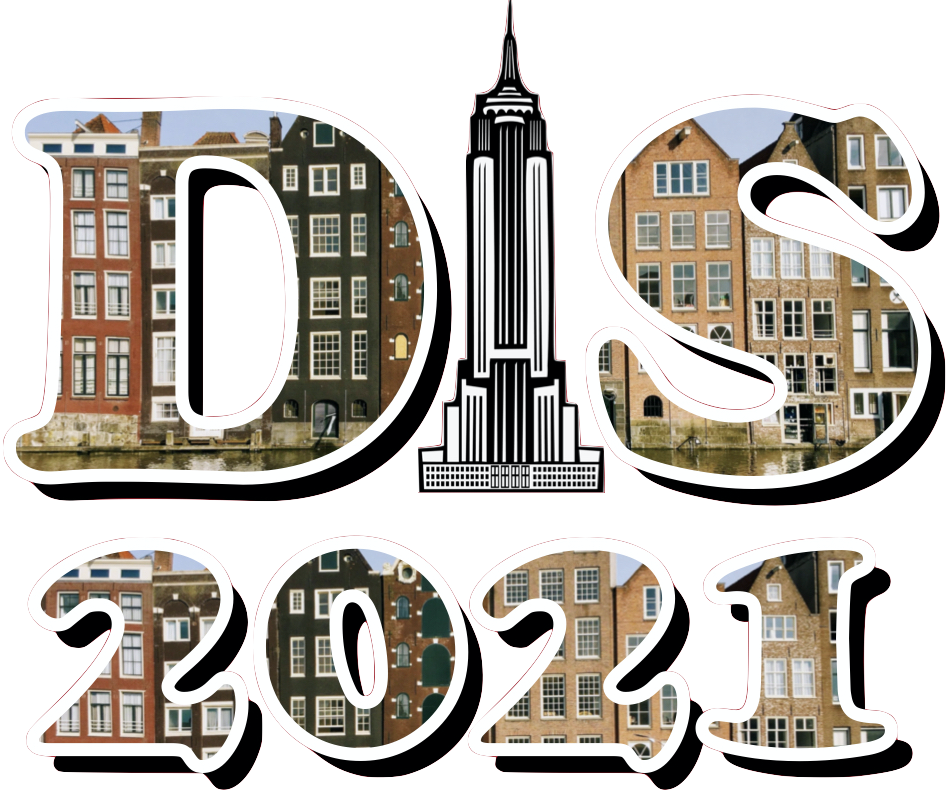}
  \end{minipage}
  &
  \begin{minipage}{0.75\textwidth}
    \begin{center}
    {\it Proceedings for the XXVIII International Workshop\\ on Deep-Inelastic Scattering and
Related Subjects,}\\
    {\it Stony Brook University, New York, USA, 12-16 April 2021} \\
    \doi{10.21468/SciPostPhysProc.?}\\
    \end{center}
  \end{minipage}
\end{tabular}
}
\end{center}

\section*{Abstract}
{\bf
  The impact parameter dependence of color charge correlators in the
  proton is obtained from the light front formalism in light cone
  gauge.  We include NLO corrections due to the $|qqqg\rangle$ Fock
  state via light-cone perturbation theory. Near the center of the
  proton, the $b$-dependence of the correlations is very different
  from a ``transverse profile function''. The resulting $t$-dependence
  of exclusive $J/\Psi$ photoproduction transitions from exponential
  to power law at $|t| \approx 1$~GeV$^2$.  This prediction could be
  tested at upcoming DIS facilities or in nucleus-proton
  ultraperipheral collisions (UPCs).  }


\section{Introduction}
\label{sec:intro}
The Hamiltonian light front formalism~\cite{Brodsky:1997de} in light
cone gauge provides essential insight into correlations of color
charges in the proton~\cite{Dumitru:2018vpr}. These can be expressed
as matrix elements of nonperturbative boost-invariant light cone
Fock-space wave functions of the QCD Hamiltonian, and related to
physical observables such as the exclusive final states measured in
DIS experiments. For example, at leading order the light-cone gauge
color charge correlator is related to the average quark transverse
momentum vector and to the Sivers
asymmetry~\cite{Burkardt:2003yg}. Furthermore, in the mixed transverse
momentum -- transverse coordinate space representation, the color
charge correlator can be related to the Wigner
distribution~\cite{Lorce:2011kd,Belitsky:2003nz,Ji:2003ak}, and to
various other generalized parton distribution functions. This detailed
information about the partonic structure of the proton can be accessed
experimentally e.g.\ in exclusive dijet or meson production, or in
vector meson -- lepton azimuthal correlations in DIS, as recently
argued in
refs.~\cite{Hatta:2016dxp,Hatta:2017cte,Mantysaari:2020lhf,Mantysaari:2019csc}.

The notion of color charge density fluctuations in the transverse
impact parameter plane emerges naturally in high-energy (small-$x$)
scattering. The projectile charge traverses without recoil the (color)
field produced coherently by all ``valence'' charges in the target,
and the scattering amplitude follows from a correlator of path ordered
exponentials of that field~\cite{Mueller:2001fv}. The scale separation
in soft coherent fields sourced by ``frozen'' valence charges was
introduced by McLerran and Venugopalan (MV) in
Refs.~\cite{McLerran:1993ni,McLerran:1993ka}. Their model was devised
for a very large nucleus and describes Gaussian fluctuations of
classical color charge densities. However, when the density of valence
charges in the target is not very large, one would rather take the
two-dimensional color charge density as an operator acting on the
light-cone wave function of the target~\cite{Dumitru:2018vpr}.

In this contribution we focus on the non-trivial impact parameter
dependence of color charge correlations in the proton, from which we
predict a non-exponential, power-law tail of $\mathrm{d}\sigma/\mathrm{d}t$ for
high-energy exclusive (coherent) $\mathrm{J}/\Psi$ photoproduction at large
transverse momentum transfer $|t| \ge 1$~GeV$^2$.

\section{The color charge correlator and coherent $J/\Psi$ photoproduction at high momentum transfer}

The central object of consideration is the two-point
color charge correlator
\begin{equation}
\label{eq:g2def}
\langle \rho^a(\vec q_1)\, \rho^b(\vec q_2) \rangle \equiv \delta ^{ab}\, g^2
G_2(\vec q_1,\vec q_2)~.
\end{equation}
The notation $\langle\cdots\rangle$ denotes an expectation value between proton
states, $\langle K|$ and $|P\rangle$, stripped of the delta-functions for
conservation of transverse and light-cone momentum.

The insertion of the charge operators $\rho^a(\vec q_1),\rho^b(\vec
q_2)$ between the incoming and scattered proton states corresponds to
the attachment of two static gluon probes (with amputated propagators)
to the color charges in the proton, in all possible ways. The two
static gluons that probe the proton structure carry transverse momenta
$\vec q_1$ and $\vec q_2$, and the total momentum transfer to the
proton is
\begin{equation}
    \vec K - \vec P = -(\vec q_1  + \vec q_2)~.
\end{equation}
From here on we choose $\vec P=0$ for the incoming proton.

\begin{figure}[tb]
  \begin{center}
    \includegraphics[width=0.3\textwidth]{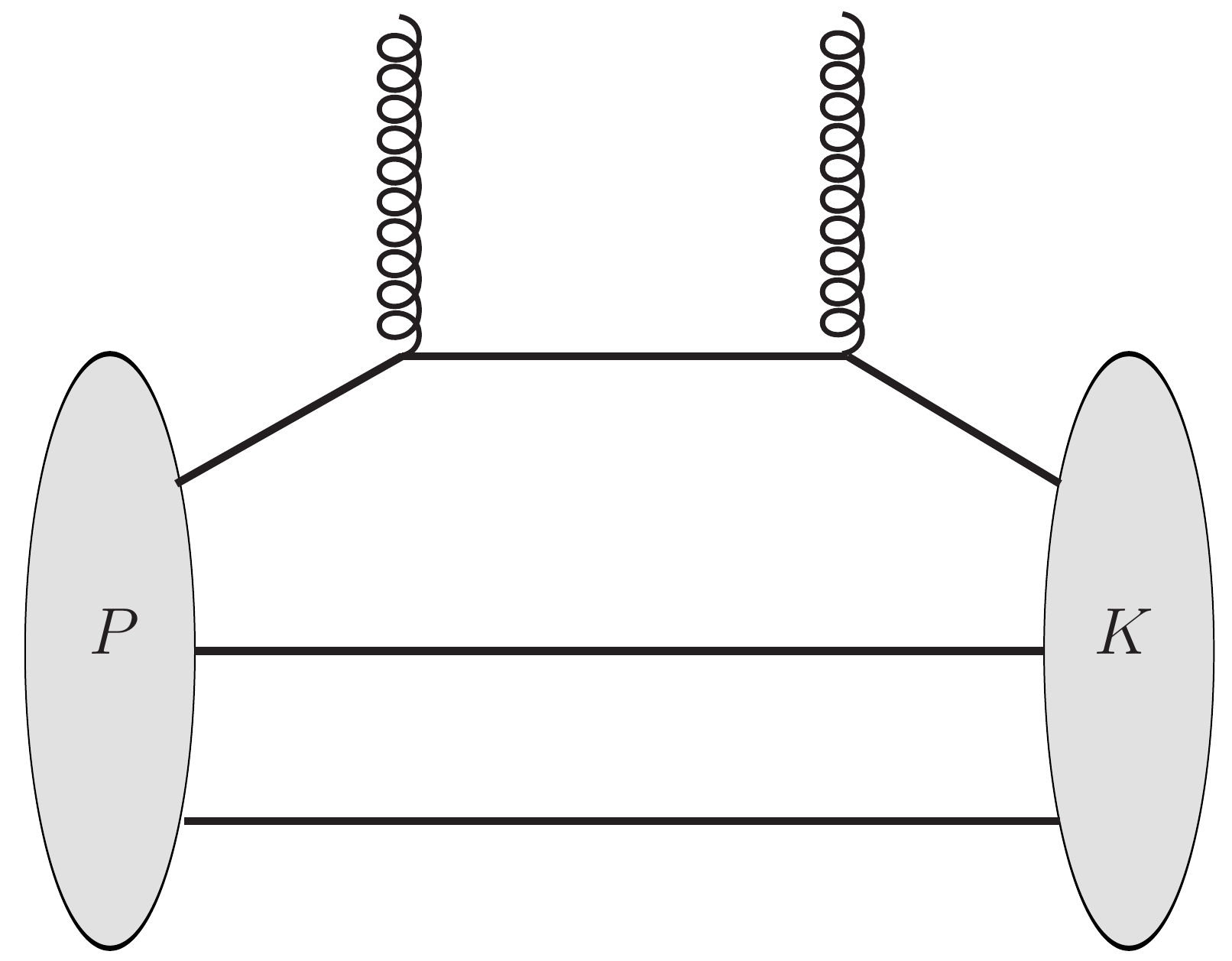}
    ~~~~~
    \includegraphics[width=0.3\textwidth]{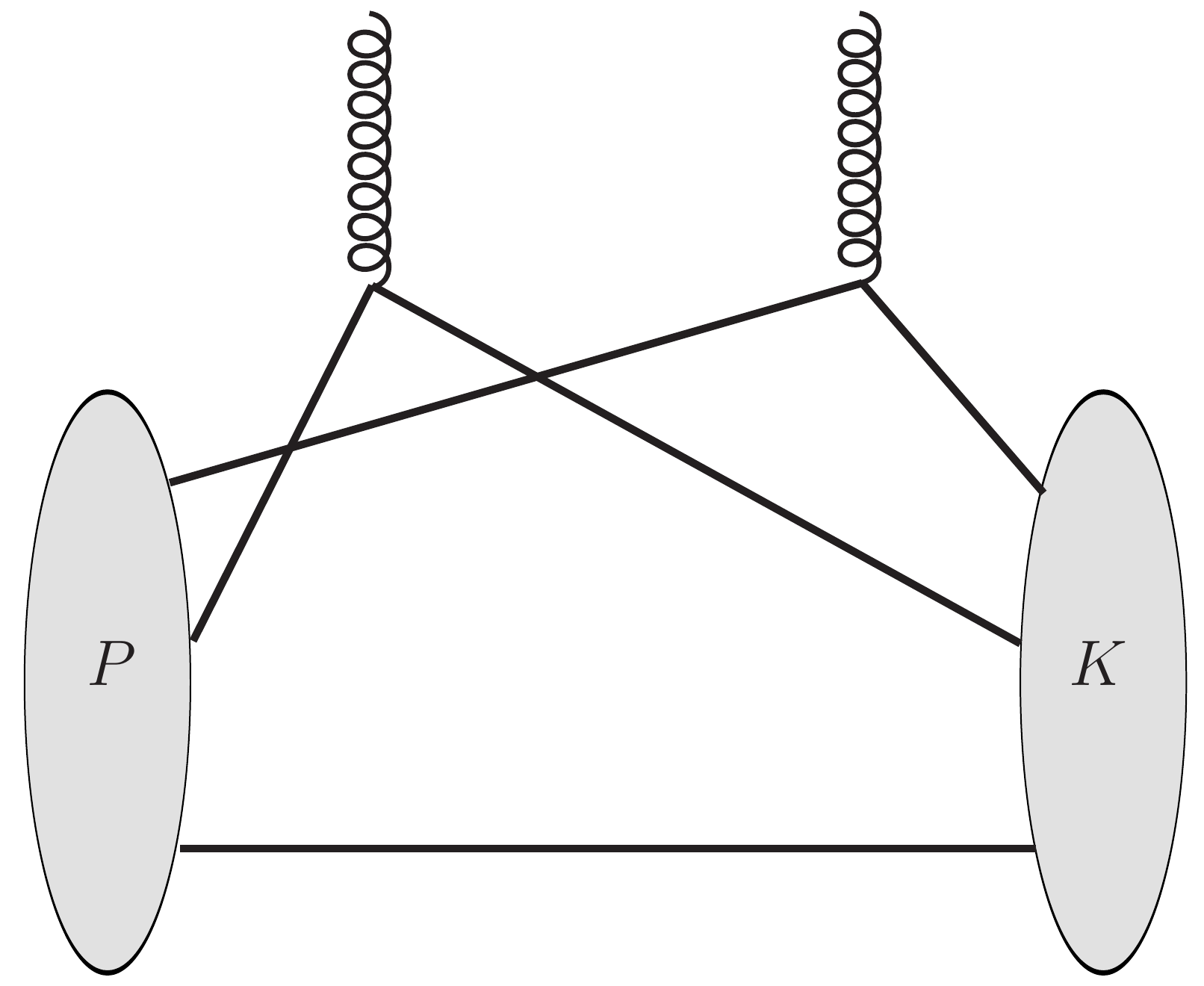}
  \end{center}
\caption{Handbag (left) and cat's ears (right) diagrams at LO.
The former (latter) dominates for small (large) momentum transfer $|t|$.}
\label{fig:Handbag-CatsEars}
\end{figure}
Figure~\ref{fig:Handbag-CatsEars} shows ``handbag'' and
``cat's ears'' diagrams at leading order (LO) ${\cal O}(g^2)$. The
former are represented by one-body operators so that the entire
momentum transfer $\vec K$ to the proton flows into a single valence
quark line. Therefore, for $\vec q_1\to-\vec q_2$ (i.e.\ $\vec K \to
0$) there is maximal wave function overlap for the LO handbag
diagram. This diagram is proportional to the electromagnetic (Dirac) form
factor, i.e.\ to the distribution $\langle\rho(\vec q)\rangle$ of
electric charge in the proton. See, for example, Eqs.~(9,10)
  in ref.~\cite{Dumitru:2019qec}.

For large momentum transfer $|t| \simeq \vec K^2$, on the other hand,
wave function overlap in the handbag diagram is highly
suppressed. Here, the overlap of incoming and scattered proton is much
greater for the cat's ears diagram because the momentum transfer is
shared by two (or even three, at NLO) valence quarks. Since $\vec K$
is the Fourier conjugate to the two-dimensional (2D) transverse
coordinate vector (impact parameter) $\vec b$, it follows that color
charge correlators near the center of the proton are dominated by
diagrams where the momentum transfer is shared by multiple partons
($n$-body operators).\\

NLO corrections due to the emission of a perturbative gluon by one of
the quarks (plus the corresponding virtual corrections) have been
computed in ref.~\cite{Dumitru:2020gla}. These are suppressed by an
additional factor of $\alpha_s$ but enhanced by a logarithm of the
minimal light-cone momentum fraction $x$ of the
gluon~\cite{Lipatov:1976zz, Kuraev:1977fs, Balitsky:1978ic}.  Diagrams
where a quark exchanges a gluon with itself across $\langle K| \to
|P\rangle$ also involve a DGLAP~\cite{Gribov:1972ri, Altarelli:1977zs,
  Dokshitzer:1977sg} collinear logarithm. Numerically, the one-gluon
emission corrections to the color charge correlator $G_2$ are small at
$x\simeq0.1$ but turn into the dominant contribution at
$x\simeq0.01$~\cite{Dumitru:2021tvw}.\\

To show the behavior of $G_2$ as a function of impact parameter we perform
a Fourier transform w.r.t.\ the momentum transfer,
\begin{equation}
  \label{eq:G2_q12_b}
  G_2(\vec q_{12},\vec b) = \int \frac{\mathrm{d}^2 \vec K}{(2\pi)^2} \,
  e^{-i\vec b\cdot \vec K}\,
\, G_2\!\left(\frac{\vec q_{12}-\vec K}{2}, -\frac{\vec q_{12}+\vec
  K}{2} \right)~.
\end{equation}
Here $\vec q_{12} = \vec q_1 - \vec q_2$ denotes the relative
transverse momentum of the two gluon probes, the Fourier conjugate to
the transverse distance $\vec r$ between the two gluons. For $\vec
q_{12}=0$, the integral of $G_2$ over the transverse impact parameter
plane vanishes,
\begin{equation}
\int \mathrm{d}^2 \vec b \,\, G_2(\vec q_{12}=0,\vec b) = 0~.
\label{eq:G2-sum_rule}
\end{equation}
This follows from the fact that $G_2(\vec q_1,\vec q_2)$ satisfies a Ward
identity and vanishes when either $\vec q_i\to0$~\cite{Bartels:1999aw,
  Ewerz:2001fb, Dumitru:2020fdh, Dumitru:2020gla}.

\begin{figure}[htb]
\centering
\includegraphics[width=0.49\textwidth]{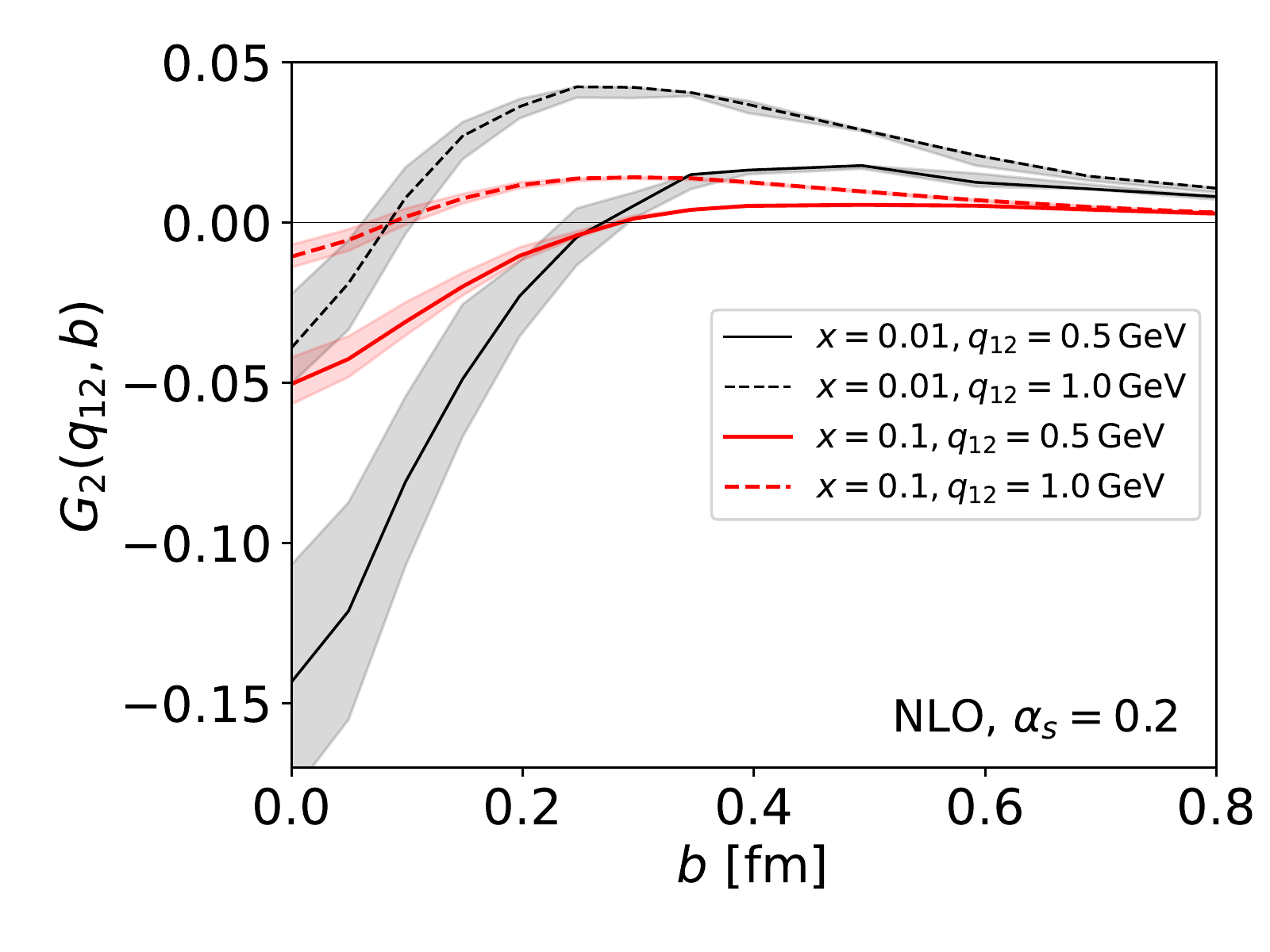}
\includegraphics[width=0.49\textwidth]{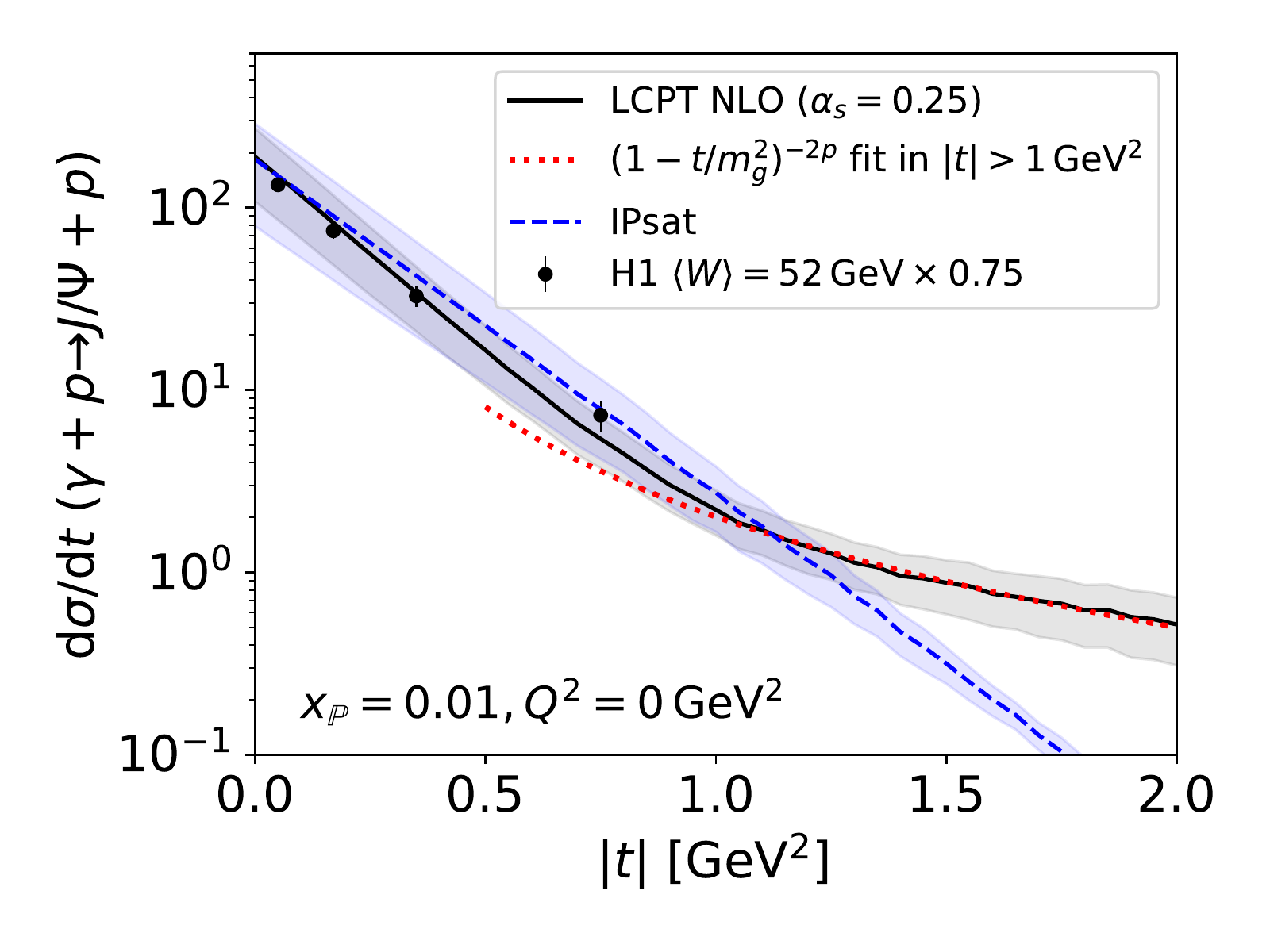}
  \caption{Left: Color charge correlator $G_2(\vec q_{12}, \vec b)$ as a
    function of impact parameter for various relative gluon transverse
    momenta $\vec q_{12}$~\cite{Dumitru:2021tvw}. The bands indicate
    the variation with the collinear cutoff, over the
    range $m=0.1 \dots 0.4$~GeV.\\
  Right: cross section for exclusive $J/\Psi$ photoproduction versus
  squared transverse momentum transfer. We compare predictions of our
  light-cone perturbation theory approach and of the IPsat model at
  $x_\mathbb{P}=0.01$ (or $W\approx 30$ GeV).  The bands reflect the
  uncertainty of the $\mathrm{J}/\Psi$ wave
  function~\cite{Lappi:2020ufv}.  For reference, we also show scaled
  data by the H1 collaboration~\cite{Alexa:2013xxa} taken at higher
  energy.}
\label{fig:mixed_space_finite_q12}
\end{figure}
Figure~\ref{fig:mixed_space_finite_q12} shows $G_2$ versus $b$.  Near
the center of the proton we find that $G_2<0$, i.e.\ ``repulsive''
two-body correlations dominate here. Also, comparing $x=0.1$ to
$x=0.01$ we note that the NLO correction mainly affects $G_2$ at small
$b$ to strongly boost these negative two-body correlations, more so
for smaller $q_{12}$. However, hand-bag type contributions become more
prominent with increasing $q_{12}$ or $b$ (the standard GPD
limit). Generically, the large-$b$ tails of the two-point correlator
$G_2$ exhibit a fall-off that resembles a transverse profile function.

We may compare the above result from LCPT to the widely used IPsat
model~\cite{Kowalski:2003hm,Mantysaari:2018nng}. Here, the color charge correlator is assumed
to factorize into a function of $\vec q_{12}$ (or of $\vec r$, in coordinate
space representation) times a Gaussian spatial profile function
\begin{equation}  \label{eq:Tp(b)}
  T_p(\vec b) = \frac{1}{2\pi B}\, e^{-b^2/2B}~.
\end{equation}
A similar factorized form of $G_2$ was proposed in
ref.~\cite{Kovner:2018fxj}. In such models the correlator of course does not
change sign as a function of $b$. Also, it does not satisfy the sum
rule eq.~(\ref{eq:G2-sum_rule}).\\

From the color charge correlator we can compute the scattering amplitude
of a quark-antiquark dipole as a function of its size and of impact
parameter~\cite{Dumitru:2018vpr}:
\begin{eqnarray}
    N(\vec r,\vec b) &=& -g^4 C_F
  \int \frac{\mathrm{d}^2 \vec K \mathrm{d}^2 \vec q}{(2\pi)^4}
  \frac{\cos\left(\Vec b \cdot \vec K\right)}{(\vec q - \frac{1}{2}\vec K)^2\,\,
    (\vec q + \frac{1}{2} \vec K)^2} \nonumber\\
& &  \times \left( \cos(\vec r \cdot \vec q) -
  \cos\left(\frac{\vec r \cdot \vec K}{2} \right) \!\! \right)
  \,\,
  G_2\left(\vec q -\frac{1}{2}\vec K, -\vec q - \frac{1}{2} \vec K\right)~. 
\end{eqnarray}
This expression applies in the two-gluon exchange approximation in the
regime of weak scattering, $N(\vec r,\vec b) \ll 1$ since it does not
resum the Glauber-Mueller multiple scattering series. To perform such
resummation, the color charge correlator would have to be transformed
from light cone to covariant gauge.

The dipole scattering amplitude exhibits an interesting dependence on
the azimuthal angle between $\vec r$ and $\vec b$ which is discussed in
more detail in ref.~\cite{Dumitru:2021tvw}.
\\


We now employ our color charge correlator $G_2(\vec q_1,
\vec q_2)$ obtained from light-cone perturbation theory (LCPT) to
compute the cross section for $\gamma + p \to J/\Psi + p$ as a
function of transverse momentum transfer. Our main interest is in the
regime of relatively high momentum transfer where, as explained above,
the quark-antiquark dipole in the photon predominantly scatters from
multiple partons in the proton.

The computation of $\mathrm{d}\sigma/\mathrm{d}t$ from the dipole
scattering amplitude is described in detail in the literature, see for
example refs.~\cite{Kowalski:2006hc,Mantysaari:2016jaz,Lappi:2020ufv}
(and ref.~\cite{Mantysaari:2021ryb} for a first next-to-leading order
calculation presented at this Workshop).  Our result at
$x_\mathbb{P}=0.01$ is shown in
Fig.~\ref{fig:mixed_space_finite_q12}. (For reference, we also show
HERA data taken at higher energy, $x_\mathbb{P}\approx 0.0035$ shifted
to match the normalization of our calculation.  However, our main
focus here is on the $t$-dependence.)

The IPsat model leads to an exponential drop off with $|t|$ which of
course follows from the parameterized $b$-dependence
eq.~(\ref{eq:Tp(b)}).  The proton structure that emerges from our LCPT
computation exhibits a similar exponential fall-off up to $|t| \approx
0.8$~GeV$^2$.  However, this turns into a power-law dependence at
greater momentum transfer. The cross section at large $|t|$ probes
color charge correlations at small $b$, near the center of the proton.
As shown in the previous section, there our correlation function
$G_2(\vec b, \vec q_{12})$ is very different from a ``profile function''.

A description of $\mathrm{d}\sigma/\mathrm{d}t$ in terms of a dipole form factor has
been proposed previously~\cite{Frankfurt:2002ka} in order to account
for the non-exponential slope at small {\em transverse momentum}
transfer, in particular for low energies near the kinematic production
threshold. Instead, here we consider the high-energy limit far above
threshold and high $|t| \simeq K_T^2 > 1$~GeV$^2$, where the cross
section is {\em not} expressed in terms of a form factor but in terms
of a two-point correlation function of color charge densities in the
proton, as explained above.

\section{Conclusion}
In conclusion, we suggest that the impact parameter dependence of
color charge correlations in the proton (at moderately small $x$) is
very different from a ``transverse profile function'', especially at
small $b$. In particular, the two-point correlation function is
negative at $b\to 0$, and changes sign at intermediate impact
parameters.

The non-trivial impact parameter dependence of color charge correlation
functions leads to a $t$-dependence of the cross section for exclusive
$J/\Psi$ production which exhibits an exponential fall-off with $|t|$
at intermediate momentum transfer, which then changes into a power-law
(approximately dipole-like) fall-off beyond $|t|\approx1$~GeV$^2$.


\paragraph{Funding information}
This work was supported by the Academy of Finland, projects 314764
(H.M) and 1322507 (R.P). H.M.\ is supported under the European Union’s
Horizon 2020 research and innovation programme STRONG-2020 project
(grant agreement no.\ 824093), and R.P.\ by the European Research
Council grant agreement no.\ 725369. A.D.\ thanks the US Department of
Energy, Office of Nuclear Physics, for support via Grant DE-SC0002307.



\bibliography{refs.bib}

\nolinenumbers

\end{document}